\documentclass[11pt]{article}
\usepackage{fancyhdr}
\def\eqalign#1{\null\,\vcenter{\openup\jot
  \ialign{\strut\hfil$\displaystyle{##}$&$\displaystyle{{}##}$\hfil
      \crcr#1\crcr}}\,}

\usepackage{color}
\usepackage{hyperref}
\usepackage{graphicx}
\usepackage{float}

\def\iniz{\setcounter{equation}{0}{%
\rhead{\thepage}\lhead{{{{\small\bf\thesection:}
\small \ \SEC\ \  \tiny\today}}}}}

\def\inizA{\setcounter{equation}{0}{%
\rhead{\thepage}\lhead{{{{\small\bf\thesection:}\ \SEC\ \  \tiny\today}}}}
\renewcommand{\theequation}{\Alph{section}.\arabic{equation}}
}
\let\a=\alpha   \let\g=\gamma  \let\d=\delta \let\e=\varepsilon
\let\z=\zeta  \let\h=\eta   \let\th=\vartheta  \let\l=\lambda
\let\m=\mu    \let\n=\nu    \let\x=\xi     \let\p=\pi    \let\r=\rho
\let\s=\sigma \let\t=\tau   \let\f=\varphi 
\let\ch=\chi     
\let\G=\Gamma \let\D=\Delta  \let\Th=\Theta\let\L=\Lambda \let\X=\Xi
     \let\F=\Phi    \let\Ps=\Psi
\let\O=\Omega 

\def\V#1{{\bf#1}}
\def\*{\vskip 3mm}\def\0{\noindent}
\def\be{\begin{equation}}
\def\ee{\end{equation}}
\def\bea{\begin{eqnarray}}
\def\eea{\end{eqnarray}}
\renewcommand{\theequation}{\arabic{section}.\arabic{equation}}

\def\NN{{\cal N}}\def\FF{{\cal F}}
\let\fra=\frac
\def\ie{{\it i.e.}\ }

\def\lis#1{\overline#1}
\def\defi{{\buildrel def\over=}}
\def\media#1{{\Blangle\,#1\,\Brangle}}
\def\bra#1{{\Blangle#1\Bvert}}\def\ket#1{{\Bvert#1\Brangle}}
\def\braket#1#2{\Blangle#1\Bvert#2\Brangle}
\def\otto{\,{\kern-1.truept\leftarrow\kern-5.truept\to\kern-1.truept}\,}
\def\wt#1{{\widetilde#1}}
\def\wh#1{{\widehat#1}}
\def\tende#1{\,\vtop{\ialign{##\crcr\rightarrowfill\crcr
 \noalign{\kern-1pt\nointerlineskip} \hskip3.pt${\scriptstyle
 #1}$\hskip3.pt\crcr}}\,}
\def\ie{{\it i.e.\ }}
\def\FF{{\mathcal F}}

\newdimen\xshift \newdimen\xwidth \newdimen\yshift \newdimen\ywidth
\def\fline{\hbox to\hsize}
\def\ins#1#2#3{\vbox to0pt{\kern-#2pt\hbox{\kern#1pt #3}\vss}\nointerlineskip}

\def\eqfig#1#2#3#4#5{
\par\xwidth=#1pt \xshift=\hsize \advance\xshift
by -\xwidth \divide\xshift by 2
\yshift=#2pt \divide\yshift by 2%
\fline{\hglue\xshift \vbox to #2pt{\vfil
#3 \includegraphics{#4.eps}
}\hfill\raise\yshift\hbox{#5}}}

\def\Bth  {{\mbox{\boldmath$ \vartheta$}}}

\def\Bs   {{\mbox{\boldmath$ \sigma$}}}

\def\Brangle {{\mbox{\boldmath$ \rangle$}}}
\def\Blangle {{\mbox{\boldmath$ \langle$}}}
\def\Bvert{{\mbox{\boldmath$|$}}}

\def\Eq#1{\label{#1}}
\def\equ#1{(\ref{#1})}

\begin{document}

\kern-1cm
{\color{red}
\centerline{\bf\large Random matrices and Lyapunov coefficients
  regularity}} \*

{\color{blue}%
\centerline{\tt G.Gallavotti}} 
{\color{blue}\centerline{\small INFN-Roma1 and Rutgers
  University}} {\color{red}\centerline{\today} }

\* \0{\bf Abstract: \it Analyticity and other properties of the largest or
  smallest Lyapunov exponent of a product of real matrices with a ``cone
  property'' are studied as functions of the matrices entries, as long as
  they vary without destroying the cone property. The result is applied to
  stability directions, Lyapunov coefficients and Lyapunov exponents of a
  class of products of random matrices and to dynamical systems. The
  results are not new and the method is the main point of this work: it is
  is based on the classical theory of the Mayer series in Statistical
  Mechanics of rarefied gases.}

\* \def\SEC{Introduction and paradigm}
\section{\SEC}\iniz\label{sec1}

Regularity of the Lyapunov exponents of products of random matrices has
been studied thoroughly in \cite{Ru979}. The study dealt with various
aspects and consequences of the following {\it cone property} (as it will
be called here):

\* \0{\bf Definition 1}: 
\label{D1} {\it A sequence $\{T_j\}_{-\infty}^{\infty}=
\ldots,T_0,T_1,T_2,\ldots$ of $d\times d$ real
  invertible matrices ``has the $(\G,{\G\,}')$-cone property'' if there are
  proper, closed convex cones $\G,{\G\,}'\subset R^d$, with apex at the
  origin $O$ and ${\G\,}'\subset \G$, such that $T_j \G\subset {\G\,}'$ and
  ${\G\,}'/O$ is contained in the interior $\G^0$ of $\G$.}  \*

\0It will be convenient to imagine the matrix $T_j$ attached to the point
$j$ of a lattice to which is also attached the linear space $E_n=R^d$ on
which $T_j$ acts transforming a vector $v\in E_j$ into $T_jv$ which is
regarded as an element of $E_{j-1}$: $T_j E_j=E_{j-1}$.

The proofs in \cite{Ru979} are based on the implicit function theorem and
on the results in \cite{FK960}.  Here the aim is to obtain most of the
results in \cite{Ru979}, relative to the finite dimensional case and
product of matrices, with a different and self contained technique.  The
``cone property'' importance for studies beyond the Lyapunov exponents,
like decay of correlations in smooth and non smooth dynamical systems, has
been developed in \cite{Li995}: the new techniques of the latter work (and
in the many stemming out of it) are also quite different from the ones
presented here which rely on the often vituperated cluster expansion, used
here in the form originated in\cite{Ru964}, which nevertheless remains one
of the simplest and most powerful techniques in Statistical Mechanics and
Renormalization Theory.

To present the main idea of this work imagine the $T_n$ diagonalizable; but
the results will cover the general case.  Then
$T_n$ will be written:
\be T_{n}=\sum_{\s=0}^{d-1} \l_{n,\s}\, {\ket{\s,n}\bra{\s,n}}
\Eq{e1.1}\ee
where $\l_{n,\s}$ are the eigenvalues of $T_{n}$ and the vectors
$\ket{n,\s},\bra{n,\s}$ are the corresponding right and left eigenvectors
which will be supposed normalized to
$\braket{n\s}{n,\s'}\equiv\d_{\s\s'}$. The eigenvalues will be labeled by
decreasing modulus $|\l_{n,0}|\ge|\l_{n,1}|\ge\ldots\ge |\l_{n,d-1}|$ and
$\l_{n,\s}\ne0$ (invertibility condition).

Notice that largest Lyapunov exponent vectors $h(n)$, assumed existing,
have to be inside the cone $\G'$ and should satisfy $\L(n,p)
h(n)=T_{n}\cdots$ $ T_{p-1} h(p)$ for $p\ge n$ and $\L(n,p)$ suitable.
Therefore, suitably fixing a convenient normalization for $h(n)$,
it should be obtainable as a limit as $N\to\infty$ of $H(n,N)\defi\,
T_{n}\cdots T_N \, \, v,\,\, v\in {\G\,}'$ which can be written as
\be \eqalign{&
H(n,N)=\sum_{\s_{n},\ldots,\s_{N}} 
\ket{n,\s_{n}}\cdot
\Big(\prod_{j=n}^{N}\l_{j,\s_j}\Big)\prod_{j=n}^{N}
\braket{j,\s_j}{j+1,\s_{j+1}}
\cr}
\Eq{e1.2}
\ee 
with 
$\ket{N+1,\s_{N+1}}\equiv v$ (here $\s_{N+1}$ is just a label and not an index
of summation, being only used for uniformity of notation). 

The representation in Eq.\equ{e1.2} suggests an alternative approach to the
analysis of such products of matrices, directly inspired by the methods of
{\it $1$-dimensional} statistical mechanics of spin systems.  

To bring it to a more familiar form: consider intervals of integers
$J=[h,h']=(h,h+1,\ldots,h')$ with $1\le h\le h'\le N$ and, on them, {\it spin
  configurations} $\Bs_J=(\s_h,\ldots,\s_{h'})$ with $\s_j=1,\ldots, d-1$
(\ie $\s_j\ne0$, see Eq.\equ{e1.1}).

Call $Y=(J,\Bs_J)$ a ``{\it polymer}'' with base $J$ and structure $\Bs_J$
(if $|J|=\ell$; there are $\ell^{d-1}$ polymers with base $J$): then
Eq.\equ{e1.2} can be interpreted as an expectation value evaluated
in an ensemble of polymers as follows.

A ``{\it configuration}'' of polymers in $[1,N]$ will be $\V
Y=(J_1,\Bs_{J_1},\ldots,J_s,\Bs_{J_s})$ with $J_1<J_2<\ldots <J_s$, $s>0$:
\ie $\V Y$ is a configuration of non overlapping polymers (``hard core
polymers'' of size $|J_i|\ge1$).  With each polymer $(J,\Bs_J)$ associate
an ``activity'' $I(J,\Bs_J)$ ($I$ might even be complex).


\eqfig{100}{40}
{
\ins{-3}{33}{$0$}
\ins{8}{33}{$0$}
\ins{18}{33}{$1$}
\ins{28}{33}{$1$}
\ins{38}{33}{$0$}
\ins{47}{33}{$0$}
\ins{56}{33}{$0$}
\ins{67}{33}{$1$}
\ins{76}{33}{$2$}
\ins{87}{33}{$1$}
\ins{98}{33}{$0$}
}
{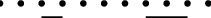}{}

\kern-4mm
\0{\small Fig.1: Spin configuration for $3\times3$ matrices containing
two polymers, of sizes $2,3$.}
\*

A formal probability distribution (``ensemble'') on the $\V Y$'s is
obtained by attributing a {\it weight} $\z(\V Y)\defi\prod_{i=1}^s
I(J_i,\Bs_i)$ equal to the product of the activities; the empty
configuration is given weight $1$. The ensemble thus defined is formal as
$I(J,\Bs_J)\ge0$ is not required .

After some meditation, it is recognized that Eq.\equ{e1.2}, can be
rewritten imagining the sites with $\s_i=0$ as ``empty sites'' in polymer
configurations, and it can be cast in the form
\be H(n,N)=\lis \L(n,N)\sum_{\s_n,\ldots,\s_{N}}
\sum_{s\ge0}\sum_{J_1<\ldots<J_s} \ket{n,\s_{n}}\,\frac{\prod_{i=1}^s
  I(J_i,\Bs_{J_i})}{\O(n,N)}\, \Eq{e1.3}\ee
where $\lis \L(n,N),\O(n,N)$ are normalization factors and
\be\eqalign{
&I(J,\Bs_J)\defi
\Big(\prod_{j=h}^{h'}\frac{\l_{j,\s_j}}{\l_{j,0}}\Big)
\Big(\prod_{j=h}^{h'-1}
\frac{\braket{{j,\s_j}}{{j+1,\s_{j+1}}}}{\braket{{j,0}}{{j+1,0}}}\Big)
\cr&\kern1.8cm\cdot 
\Big(\frac{\braket{{h-1,0}}{{h,\s_{h}}}}
{\braket{{h-1,0}}{{h,0}}}\Big)^{\d_{h>n}}
\Big(\frac{\braket{{h',\s_{h'}}}{{h'+1,0}}}{\braket{{h',0}}{{h'+1,0}}}
\Big)
\cr
&\O(n,N)\defi\sum_{s\ge0}\sum_{
{J_1<\ldots<J_s}\atop{\Bs_{J_1},\ldots,\Bs_{J_s}}}
\prod_{i=1}^s
I(J_i,\Bs_{J_i})
\cr
&\lis \L(n,N)\defi \O(n,N)\prod_{j=n}^N\Big( \l_{j,0}\
\braket{{j,0}}{{j+1,0}}\,\Big)
\cr}\Eq{e1.4}\ee
where $\ket{N+1,\s}$ has to be interpreted, {\it instead}, as $v$, see
Eq.\equ{e1.2}, and, for $s>0$, $J_1<\ldots< J_s$ are consecutive intervals
in $[n,N]$ {\it not empty and disjoint}; $s=0$ contributes $1$ to
$\O(n,N)$.

Eq.\equ{e1.3} maps the problem of studying $H(n,N)$ into the study of
$\lis\L(n,N)$ times a {\it formal average} in what is known in statistical
mechanics as a {\it Fisher model}, \cite{Fi967}.  The latter is well
known, since \cite{Fi967}, as a machine for examples and counterexamples in
statistical mechanics and in dynamical systems, for some applications see
\cite{Ga976}, \cite[Def. D7.3.1]{GBG004}).  \*

Here Eq.\equ{e1.3} will be the starting point to obtain, with an
alternative method, the following theorems, special cases
of results already in \cite{Ru979}:
\*

\0{\bf Theorem 1:} {\it (1) Let $T_n$ be a sequence as in definition
  \ref{D1} (hence with the $(\G,{\G\,}')$-cone property) with
  $||T_n||<B_0$ for some $B_0>0$. Then given any sequence $\{v_j\}$
  of unit vectors in ${\G\,}'$, the limits
\be \eqalign{
h(n)=&\lim_{N\to\infty} \frac{T_{n}T_{n+1}\cdots T_{N-1}\, v_N}
{\lis\L_{v_N}(n,N)}=\lim_{N\to\infty}
\frac{{H_{n,N}}{ v_N}}{\lis\L_{v_N}(n,N)}
\cr
\L(n,p)=&\lim_{N\to\infty} 
\frac{\lis\L_{v_N}(p,N)}{\lis\L_{v_N}(n,N)}>0,
\cr}\Eq{e1.5}\ee
exist $\forall 1\le n\le p$ and are independent of the sequence
$\{v_j\}_{j\ge1}$.
\\ 
(2) The limits are (real) analytic functions  
of each of the matrices entries, as long 
as their variations are small enough.  
\\
(3) The vectors $h(n)$ are ``eigenvectors'' for the product of the
inverse matrices $T_j^{-1}$, \ie there are ``Lyapunov
coefficients'', $\L(n,p)$ such that:
\be T^{-1}_{p-1}\ldots T^{-1}_{n}\, h(n)\,=\,\L(n,p)\, h(p),\qquad p> n
\Eq{e1.6}\ee
(4) There is $B$ such that $B^{-1}\le ||h(n)|| \le B$, $\L(n,p)>0$, and the
upper and lower limits of $\frac1p\log \L(n,p)$, as ${p\to \infty}$,
are $n$  independent. }
\*

\0{\bf Remarks:} 
(a) The uniqueness property implies that $h(n)$ is the {\it unique} 
eigenvector (\ie invariant vector) with the largest Lyapunov exponent.
\\ 
(b) $h_n=\frac{h(n)}{||h(n)||}$ is called {\it unstable
unit ({\rm or} {\it direction}) vector} at site $n$.  \\
(c) Imagine the matrices $(T_j)_{j=1}^\infty$ be a sequence of random
variables and that their entries are distributed with a distribution $\r$
which is invariant with respect to the (left) translations, and with
samples restricted to keep the cone property, \ie with the cones $\G,{\G\,}'$
which do not depend on the choice of the matrices. Then the upper and lower
limits of $\frac1p\log |\L(n,p)|$ as $p\to\infty$ will be constant under
translation, \ie $n$-independent as a consequence of the $n$-independence
in item (4). They will be shown below to exist almost everywhere, as a
consequence of the ergodic theorem, hence if $\r$ is ergodic they will be
equal and constant with $\r$ probability $1$.
\\
(d) As a corollary a general analyticity property holds: 
\*

\0{\bf Theorem 2:} {\it Let $\{T_j\}_{-\infty}^\infty$ depend on a real
parameter $z\in[-\lis\n,\lis\n]=\D$ and admit, for each $j$, a power
series  with radius $\n$ around each point in $\D$; and, furthermore, for all
$z\in \D$ they have the cone property with respect to $z$--independent
cones $\G,\G'$. Then the vectors $h(n)$ and the Lyapunov coefficients
$\L(n,p)$ are holomorphic in $z$ for $|z-\D|<\n'$ for some $\n'$.}  \*

Theorems 1,2 are paradigms: they are the basis for similar theorems for
dynamical systems, {\it e.g.} \cite{Ru979}, and Appendix \ref{appC} where
are formulated the immediate (classical) applications of the results to
dynamical systems.
\\ (e) It shoould be stressed that the results in the theorems above are
   {\it well known}: the purpose of this note is to remark that the problem
   is naturally formulated as a problem in one dimensional Statistical
   Mechanics with short range interactions and, as such, it can be
   immediately solved by the simplest method of the theory of dilute gases,
   namely the Mayer expansion.  
\\ This becomes clear just after the estimate claimed in Eq.\equ{e2.2}
below, whose proof is completed at Eq.\equ{e2.13}.  Appendix \ref{appB}
reproduces {\it for completeness} the classic theory of the cluster
expansion in its primitive form, based on Ruelle's ``algebraic formalism''
(as subsequently elaborated in \cite{GM968} and mainly in \cite{GK971,Ca982})
following the exposition in \cite{GBG004}). Since the appendix is here only
for readers that are not already familiar with the expansion, no effort has
been devoted to obtain `''best'' (nor better) estimates: much progress has
been made on the cluster expansion and the convergence estimates can be
greatly improved, \cite{Ca982,KP986,Do996,Mi000,FP007} (I thank a referee for
suggesting the latter path through a vast literature on the subject).

The expression Eq.\equ{e1.4} suggests also to study the statistical
mechanics problem via another key technique used in one dimensional
statistical mechanics, namely the {\it transfer matrix technique}. The
approach is possible, particularly if the matrices have non negative
entries and it has been applied, \cite[p.69]{Ru979} and with attention to a
constructive approach in \cite{Pe992,Po010}, where it is shown that the
properties of the largest exponent can be studied by the transfer matrix
method and are directly related to simple spin-glass problem. The work
\cite{Po010} gives also a fast and constructive method to determine the
largest exponent within a prefixed approximation based in the transfer
matrix method.

\def\SEC{Theorems 1,2}
\section{\SEC}\iniz\label{sec2}

The essence of the proof lies in understanding the case in which the
matrices $T_j$ are diagonalizable with real eigenvectors and eigenvalues
uniformly (in $j$) pairwise separated, and possess a $(\G,{\G\,}')$-cone
property; furthermore  
  $\max_{n,\s=1,\ldots,d-1}\frac{|\l_{\s,n}|}{|\l_{0,n}|}$ $<\e_0$ with
  $\e_0$ small enough.

The more general case contemplated in theorem 1 (\ie just the cone property
and a uniform bound on $||T_n||$ is supposed) will be eventually reduced to
the latter one via the following algebraic lemmata, see proof in Appendix
\ref{appA}: \*

\0{\bf Lemma 1:} {\it Let $T$ be a (single) $d\times d$ matrix with the
  $(\G,{\G\,}')$-cone property. Then: 
\\
(a) $\frac{T^nv}{||T^n
    v||}\tende{n\to+\infty} b$ exponentially fast and $\frac{T b}{||T
    b||}=b$ for all $v\in\G$.  A corresponding property holds for the
transposed  $T^*$ with $b$ replaced by a $b^*$.
\\
(b) The eigenvalue $\l_0$ of $T$
  with maximum modulus is simple and positive, hence  there is $\g<1$
such that $\max_{\s>0}\frac{|\l_\s|}{|\l_0|}\le
  \g<1$.} \*

\0{\bf Definition 4:}\label{D4}
{\it Let $\l$ the modulus of the largest modulus eigenvalue of a matrix 
and $\l'$ the maximum  modulus of the other eigenvalues; call, here,
$\frac{\l'}\l$ the matrix ``{\it spectral gap}''.}
\*

The proof of
lemma 1 leads to, see Appendix \ref{appA}, 
\*

\0{\bf Lemma 2:} {\it Suppose that the sequence $T_1,T_2,\ldots$ satisfies
  the cone property with respect to the pair of cones $\G\supset {\G\,}'$ and
  let $T'\defi T_1\cdot T_2\cdots T_p$. Then there are constants
  $c,\a>0$ with $\a<1$ such that the spectral gap of $T'$ is $\le c
  \a^p$. Furthermore the matrix elements of $T'$ on the
  basis formed by $b$ (see lemma 1) and by $d-1$ unit vectors in the plane
  orthogonal to $b^*$ are all bounded by $c\a^p$ with respect to 
  the entry $|T^{[p]}_{0,0}|\ge\frac1c$. The $c,\a$
  depend on the inclination $\e$ and on the openings $\th,\th'$ of the
  cones.
}  \*

This implies that, if $p$ is large enough, the sequence $T'_n=T_{n
  p+1}T_{np+2}\cdots$ $ T_{(n+1)p}, n=0,1,\ldots,$ satisfies the cone
property with respect to the {\it same} pair of cones $\G\supset {\G\,}'$ and
the spectral gap of the matrices $T'_n$ can be made {\it as small as
  wished} by taking $p$ large enough.

 Hence it is sufficient to prove theorem 1 for matrices $T'_j$ with the
 above defined spectral gap $\frac{\l'}\l=\g$ as small as needed. Once the
 sequence $b'_n$ for $T'_n$ is obtained the sequence $b_n$ that has to be
 found will be obtained setting $b_{np}\defi b'_n$ and:

\be b_{np-k}=T_{np+k}\cdots T_{np} b'_{np},\quad
k=1,\ldots,p\Eq{e2.1}\ee
and all the statements in theorem 1 will, as well, follow for the sequence
$T_n$. \*

\0{\it proof of theorem 1}: Suppose {\it at first} that the matrices $T_n$
have real eigenvalues {\it with reciprocal distance}, as $n$ varies,
greater than a positive lower bound.  The $\O(n,N)$ in Eq.\equ{e1.4} can be
interpreted as the partition function of a gas of polymers represented by
(lattice) intervals $J=[h,h']\subset [1,N]$ with ``base'' $J$,
``structure'' $\Bs_J=(\s_h,\ldots,\s_{h'})$ and ``activity'' $I(J,\Bs_J)$
defined in Eq.\equ{e1.4}. Of course this is only an analogy as
$I(J,\Bs_J)$, although real numbers (under the restrictive temporary
assumption) might be $<0$.

A simple bound can be set on $|I(J,\Bs_J)|$ by remarking that the cone
property implies a $n$-independent lower bound $\frac 1\d>0$ on the scalar
products appearing in the denominators in the definition Eq.\equ{e1.4}:
actually $\frac 1\d>0$ can be chosen as a lower bound for the absolute
value of the product of any pair or unit vectors in $\G$.  If $\g$ is an
upper bound to the matrices spectral gaps, see definition 4, 
Eq.\equ{e1.4} implies. {\it also defining $\h$}:

\be 
\sum_{\Bs_J}|I(J,\Bs_J)|\le ((d-1)\d\g)^{h'-h+1}\defi\h^{h'-h+1}\Eq{e2.2}\ee

\* \0{\it Remarks:} (1) As shown by the last bound, forgetting that the
activities may be be negative and treating $I(J,\Bs_J)$ as weight
(or activity) of the polymer $(J,\Bs_J)$ in the formal probability
distribution of the polymers appearing in Eq.\equ{e1.3}, the number of $J$'s
should be very small if $\h$ (\ie $\g$) is small: the ``preferred
state'' is no $J$ at all; but the $I(J,\Bs_J)$ may be negative 
and cannot be regarded as probability weights.
\\ (2) Hence it will be necessary to evaluate the ``averages'' (like
Eq.\equ{e1.3}) with respect to the ``distribution'' in which a
configuration of polymers with bases $J_1<\ldots<J_s$ and spin structures
$\Bs_{J_1},\ldots, \Bs_{J_s}$ has weight $\prod_{i=1}^s I(J_i,\Bs_{J_i})$:
this must be done {\it algebraically}, {\it i.e.}  without profiting 
of positivity properties. It is natural to have recourse to {\it
  the cluster expansion}: which is a method that was designed, in statistical
mechanics, precisely for such tasks. \*

The general theory of the cluster expansion for polymers, see
\cite[Ch.7]{GBG004} (it is recalled from scratch, {\it for completeness},
in Appendix \ref{appB}), yields a formal expression for $\O(n,N)$ as
\be \O(n,N)=\exp{
\sum_{\V Y}\f^T(\V Y)\z(\V Y)},
\qquad \z(\V Y)=\prod_{J\in \V Y} I(J,\Bs_J)\Eq{e2.3}\ee
where the summation runs over all polymer configurations $\V Y$
consisting of $Y_1,\ldots,Y_s$, with $Y_i=(J_i,\Bs_{J_i})$, in which the
constraint of no overlap on the polymers base intervals $J_i$ (implied by
the $*$ in Eq.\equ{e1.4}) is {\it dropped} and $\f^T(\V Y)$ are suitable
(real) {\it combinatorial coefficients}, see Appendix \ref{appB},
Eq.\equ{eB.8}.

The coefficients $\f^T(\V Y)$ have the important property of being {\it
  translation invariant} under a simoultaneous translation of the polymers
in $\V Y$ and of {\it vanishing unless the intervals $J$ which are bases of
  the polymers in $\V Y$ overlap} in the sense that $\cup_{J\in\V Y}J$ is a
connected interval, called a {\it cluster} (imagine to draw the $J_i$'s as
continuous segments joining their extremes).

Abridge $\f^T(\V Y)\z(\V Y)$ into $\wh\f(\V Y)$: the sum 
$\sum_{\V Y, J_i\subset [n,N]}\wh \f^T(\V Y)$ over the clusters
$\V Y$ will be an absolutely
convergent series, summing to $\log \O(n,N)$, if it will be shown that, for
a suitable choice of $r(J)$, it is
\be\m\defi \sup_{J,\Bs_J} \frac{|\z({J,\Bs_J})|}{r({J})} 
\exp{\sum_{S,\Bs_S}^*r(S)}<1\Eq{e2.4}\ee
where the sum is over the polymers $(S,\Bs_S)$ overlapping with
$J$, \ie $S\cap J\ne\emptyset$, as
recalled in a self contained proof, in Appendix \ref{appB}. 

{\it This is a non trivial property}: certainly the overlap condition
strongly reduces the number of addends in the series for $\O(n,N)$ and
helps together with the Eq.\equ{e2.4}, which implies that $\z(\V Y)$ is
exponentially small with the size of the interval covered by the clusters
of the polymers bases in $\V Y$. However the help is not sufficient and
important combinatorial cancellations have to be taken into account: they
are exhibited through the {\it remark that the coefficients $\f^T(\V Y)\z(\V Y)$
satisfy an identity}, see Eq.\equ{eB.12}, reducible to an identity known in
Physics as {\it Kirkwood-Salsburg equations}, \cite{Ga000}.

Take $r(S)\defi\h^{\frac12|S|}$, see Eq.\equ{e2.2};
if $\h=(d-1)\d\g$ is small enough:
\be\eqalign{
\m\le& \sup_{|J|=j\ge1}\frac{\h^j}{\h^{\frac12j}} \exp{
  j\Big(\sum_{\ell=0}^\infty (\ell+1) \h^{\frac12(\ell+1)}\Big)}
=\h^{\frac12}\exp \frac{\h^{\frac12}}{(1-\h^{\frac12})^2}<1
\cr}\Eq{e2.5}
\ee
Setting, see
Eq.\equ{e1.3}, 
\be h(n,N)\defi
\sum_{s\ge0}\sum_{J_1<\ldots< J_s\atop \Bs_{J_1},\ldots,\Bs_{J_s}}
\ket{n,\s_{n}} \frac{\prod_{i=1}^s
I(J_i,\Bs_{J_i})}{\O(n,N)}\Eq{e2.6}\ee
this, as remarked in Sec.\ref{sec1}, can be interpreted as
$\sum_{\s=0}^{d-1} P_{n,\s,N}\ket{n,\s}$ with

\be P_{n,\s,N}\defi \frac{\O_\s(n,N)}{\O(n,N)}
\defi\sum_{s\ge0}\sum^{*\s}_{
{J_1<\ldots<J_s}\atop{\Bs_{J_1},\ldots,\Bs_{J_s}}}
\frac{\prod_{i=1}^s
I(J_i,\Bs_{J_i})}{\O(n,N)}\Eq{e2.7}\ee
where the $*\s$ indicates that the sum is restricted to polymers
configurations with $\s_n=\s$: \ie with $n\in J_1$ and $\s_n=\s$ if
$\s\ge1$ or with $n\not\in J_1$ if $\s_n=0$.

Hence if the activities $I(J,\Bs_J)$ were non negative $P_{n,\s,N}$ would be
the probability of finding a configuration with spin $\s$ at site $n$,
and $P_{n,0,N}+\sum_{\s=1}^{d-1} P_{n,\s,N}\equiv 1$. This relation is a
purely algebraic property and is identically satisfied (as long as the
addends are well defined).

For the $P_{n,\s,N}$ with $\s\ge1$ the cluster expansion (see
Eq.\equ{eB.15}) yields
\be P_{n,\s,N}=\sum_{J\ni n,\Bs_J,\s_n=\s}\media{D_{J,\Bs_J}\wh\f^T},\qquad
|P_{n,\s,N}|\le \frac{\h^{\frac12}}{(1-\m)(1-\h^{\frac12})}
\Eq{e2.8}\ee
where $\media{D_{J,\Bs_J}\wh\f^T}\defi \sum_{\V Y}\wh\f^T((J,\Bs_J)\cup\V
Y)$; the bound is obtained by using the mentioned overlap property that the
$\wh\f^T(\V Y)$ vanish unless the bases of the polymers in $\V Y$ form a
connected interval, see below.

The overlap property shows that the sum $\media{D_{J,\Bs_J}\wh\f^T}$ is
restricted to polymers that contain the site $n$: hence the union of the
bases of the polymers contributing terms that depend on the boundary vector
$v$ must cover the whole $[n,N]$%
\footnote{The site $n$ must be contained since $J$ must contain $n$ because
  $\s_n\ge1$; and the site $N$ must be contained as otherwise the $\f^T$
  does not depend on $v$.}
 and therefore contribute a quantity that is exponentially
small as $N\to\infty$, as mentioned above.

Explicit bounds, see Eq.\equ{eB.13}, \equ{eB.14}, if 
Eq.\equ{e2.4} holds, give $|P_{n,\s,N}|\le
\sum_{j=1}^\infty\sum_{m=0}^\infty \h^{\frac
  j2}\m^m$ where $j$ is the length of the polymer $J$ containing $n$.
Therefore $P_{n,\s,N}$ depends on the boundary condition vector $v$: but
the dependence disappears in the limit $N\to\infty$. 
Hence the limits
$P_{n,\s}$ of $ P_{n,\s,N}$ as ${N\to\infty}$ exist and
\be |P(n,\s)|\le 
\cases{\frac{\h^{\frac12}}{(1-\m)(1-\h^{\frac12})}&\  {\rm if} $\s\ge1$\cr
1+\frac{\h^{\frac12}}{(1-\m)(1+\h^{\frac12})}(d-1)&\  {\rm if} $\s=0$\cr}
\Eq{e2.9}\ee
where the bound for $\s=0$ reflects the {\it algebraic}
identity (due to the probabilistic interpretation)
$P(n,0,N)+\sum_{\s=1}^{d-1}P(n,\s,N)\equiv1$ (which holds whe\-ther or not the
activities $I(J,\Bs_J)$ are $\ge0$, provided convergence holds).

Hence the limit $h(n)\defi\lim_{N\to\infty} h(n,N)$ exists and by
Eq.\equ{e1.3} is
\be\eqalign{ h(n)=&\lim_{N\to\infty} \frac{ T_{n}\ldots
    T_N\,{v}} {\lis \L_v(n,N)}=\sum_{\s=0}^{d-1} P_{n,\s}\ket{n,\s},\qquad
  {\rm with}\cr \lis
  \L_v(n,N)\defi&\O(n,N)\prod_{j=n}^N\Big(\l_{j,0}
  \braket{j,0}{j+1,0}\Big)\cr
B^{-1}\le&||h(n)|| \le B, 
\cr} \Eq{e2.10} \ee 
with $B=(1+2d \frac{\h^{\frac12}}{(1-\m)(1-\h^{\frac12})})$, if $\h$ is
small enough so that $\m,\h<1$ (together with Eq.\equ{e2.5} this
is the condition on the spectral gap that determines the parameter
$\g$). Notice that the convergence of the cluster expansion also implies
$\O(n,N)>0$ and upper and lower bounds on $\O(n,N)$:
\be \eqalign{
|\log
&\O(n,N)|\le |\sum_{\V Y} \wh\f^T(\V Y)|\le 
\sum_{\g\subset [n,N]} \sum_{\V Y}|\wh\f^T(\g\cup\V Y)|
\cr
\le &\sum_{\g\subset [n,N]} r(\g)\sum_{m=1}^\infty I_m\le \sum_{k=1}^\infty
\frac{(N-n+1)}{1-\m}\h^{\frac12k}\le 
\frac{(N-n+1)\sqrt\h}{(1-\m)(1-\sqrt\h)},
\cr
}\Eq{e2.11}
\ee
of course not uniform in $N$. Thus exhibiting the important cancellation that
occurs in the ratios $\frac{\O_\s(n,N)}{\O(n,N)}$, above, thus
estimated uniformly in $N$.

Defining

\kern-3mm
\be \eqalign{
\L(n,p)\defi&\lim_{N\to\infty} \frac{\O(p,N)\prod_{j=p}^N\Big(\l_{j,0}
  \braket{j,0}{j+1,0}\Big)}{\O(n,N)\prod_{j=n}^N\Big(\l_{j,0}
  \braket{j,0}{j+1,0}\Big)}\cr
=&\Big(\prod_{j=n}^{p-1} \l_{j,0}\,
  \braket{j,0}{j+1,0}\Big)^{-1}\lim_{N\to\infty}\frac{\O(p,N)}{\O(n,N)}
}\Eq{e2.12}\ee
$\L$ can be evaluated again by the cluster expansion, which also allows
us to see the cancellation that shows the $v$-independence of the last
limit, as:

\be \lim_{N\to\infty}\frac{\O(p,N)}{\O(n,N)}=\exp{\sum^*_{\V Y} \wh\f^T(\V
  Y)}=\exp{\sum_{q=n}^{p-1} \F(q)}
\Eq{e2.13}\ee
where $\V Y=(J_1,\Bs_1,\ldots,J_s,\Bs_s)$ and the $*$ means that the
cluster $(\cup J_i)$ overlaps with $[n,p]$ and $\F(q)$ is the sum $\sum_{\V
  Y, J_1\ge q, J_1\ni q}\wh\f^T(\V Y)$ over all polymer configurations
which are to the right of $q$ and $q$ is the first point of $J_1$: in
Eq.\equ{e2.13} numerator and denominator are exponentials of sums of many
terms which are common (see Eq.\equ{e2.3}), hence cancel, except those
relative to $\V Y$'s with bases touching $[n,p]$. Such polymers are
independent of $v$ unless their bases touch also $N$: hence their
contributions to $\F(q)$ tend to $0$ as $N\to\infty$ at fixed $n,p$.  

The positivity of $\L$ is due to the reality of $\wt\f^T$ and to the
positivity of the scalar products and of $\l_{j,0}$ in Eq.\equ{e2.12} (by
the cone property). \*

\0{\it Remarks:} (1) It is important to stress that $\F(q)$ depends only on
the matrices $T_j$ with $j\ge q$: $\F(q)= F(T_q,T_{q+1},\ldots)$ and each
$\F(q)$ is given by an absolutely convergent series because of the bounds
on $\wh\f^T(\V Y)$ in Eq.\equ{eB.13},\equ{eB.14}. Furthermore the function
$F$ {\it is independent of $q$} (\ie it is translation invariant as a
function of the sequence of matrices), because of the translation
invariance of $\f^T(\V Y)$.
\\
(2) The uniform convergence of the series defining $\F(q)$,
Eq.\equ{e2.13}, implies that all limits points as $p\to\infty$ of
$\O(n,p)^{\frac1p}$ are $n$-independent.
\\
(3) Uniformity of the limits defining $P_{n,\s}$ also holds if the matrix
elements of the matrices $T_j$ are varied keeping the cone property
independent of the variations and their norms bounded by $B_0$: this is due
to the uniformity of the bounds on $I(J,\Bs_J)$ only depending on the
inclination $\e$ of ${\G\,}'$ in $\G$, on the opening angles $\th,\th'$
between the cones and on the bounds on the matrices norms.  \\
(4) The activities $I(J,\Bs_J)$, which so far have been real valued, are
holomorphic functions of small complex variations of the matrices entries
(since it is temporarily being supposed that the eigenvalues are real and
keep pairwise a positive minimal distance) still satisfying the same bounds
possibly with slightly different constants: hence $h(n),\L(n,p)$ are
analytic in the entries.  \\
(5) The $h(n)$ form, essentially by definition once convergence has been
established, a {\it covariant family} of vectors in the sense of theorem 1,
Eq.\equ{e1.6}, as implied by Eq.\equ{e2.10}.  In fact $h(n)$ and likewise
$\L(n,p),\O(n,N)$ are given by convergent expansions (see the series in
Eq.\equ{e2.8}, for instance) in the activities $I(J,\Bs_J)$: the latter are
given by simple algrebraic expressions, see Eq.\equ{e1.4}, which are
analytic in the matrix elements of the $T_n$.  \\
(6) If the matrix elements of the $T_j$ are chosen randomly with respect to
a distribution $\r$ on $\prod_{j=-\infty}^\infty R^d\times R^d$ which is
invariant under translations and has samples satisfying the
$(\G,{\G\,}')$-cone property the limit

\be\lim_{p\to\infty} \frac1p\log \L(n,p)=
\lim_{p\to\infty} \frac1p\Big(\sum_{j=n}^{p}\log(\l_{j,0}
  \braket{j,0}{j+1,0})+\sum_{j=n}^{p-1}\F(j)\Big)
\Eq{e2.14}\ee
exists, consequence of the ergodic theorem, because the addends in the sums
are translates to the right {\it of the same function of the random
  matrices} (\ie in the language of ergodic theory they are ``Birkhoff
averages'').  \\
(7) Furthermore $n$-independence of the limit points, remarked in
Eq.\equ{e2.14}, implies that the limits in Eq.\equ{e2.14} are constants of
motion under the right translations. Hence if $\r$ is also ergodic the
limits are constant almost everywhere and the maximum Lyapunov
exponent $\l_+$ is the integral with respect to $\r$ of $\log(\l_{1,0}
\braket{1,0}{2,0}+\F(1)$: hence it is analytic in the 
parameters on which the matrices may depend.
\*

Finally the {\it assumption that $T_j$ have spectrum consisting of pairwise
  uniformly separated eigenvalues and have a large gap has to be removed}:
lemma 2 shows that the matrices $T'_j=T_{jp+1}T_{jp+2}\cdots T_{(j+1)p}$ to
which we want to apply the analysis, see remark after Lemma 2, have a
spectral representation as $T'_j=\l'_{0,j}\,\Big(\ket{j,0}\bra{j,0} +
\sum_{\s,\s'\ge1} (\Th_j)_{\s,\s'}\ket{j,\s}\bra{j,\s'}\Big)$, where the
bases $\ket{j,\s}$,$\bra{j,\s'}$ consist of the vectors $\ket{j,0}$ and
$\bra{j,0}$ and correspondingly $d-1$ other orthogonal vectors, repectively
to $\ket{j,0}$ and $\bra{j,0}$, arbitrarily chosen. The matrix elements are
bounded above by $c\a^p, \a<1$.

Since the basis vectors $\ket{j,\s},\bra{j,\s}$ for $\s>0$ need not be
eigenvectors of $T_j$ the basis can be taken real: hence the matrix
$\Th_j$ can be taken real.

Therefore it appears that this case simply leads to more complicated
formulae in which at each site $j$ are now associated {\it two spins}
$(\s_j,\s'_j)$, instead of just $1$, and the products like
$\prod_{j=h}^{h'}\frac{\l_{j,\s_j}}{\l_{0,j}}
\frac{\braket{{j,\s_j}}{{j+1,\s_{j+1}}}}{\braket{{j,0}}{{j+1,0}}}$,
appearing in Eq.\equ{e1.4} must be replaced by the product
$\prod_{j=h}^{h'}(\Th_j)_{\s_j,\s'_j}
\frac{\braket{{j,\s_j}}{{j+1,\s'_{j+1}}}}{\braket{{j,0}}{{j+1,0}}}$.

This amounts at more values of the spins associated with each site: from
$d-1$ to $(d-1)^2+1$. There is no need to perform a full spectral
decompostion and therefore to worry about degeneracies, complex eigenvalues
and eigenvectors,
\footnote{\small {\it I.e} only the largest eigenvalue, which is separated
  from the rest of the spectrum by an $h$-independent factor $<1$ (related
  to the $\a$ in lemma 1), and the relative eigenvector are needed.}
eigenvalues crossing and the like. The only
property needed is that the $\Th_j$ be as small as necessary for the
cluster expansion and to be analytic in the matrices entries: this is
achieved, as mentioned, by taking $p$ large enough.

Theorem 1 is thus proved.  It also yields, as a corollary, theorem 2. If
the $T_h$ depend on a parameter $z$ as in the assumption of theorem 2 {\it
  and} the largest eigenvalue of $T_h$ is separated, uniformly in $h$, by a
large enough gap from the rest of the spectrum: then the property remains
valid in a complex region within a distance $0<\n'\le\n$ of the real
interval $z\in\D$.

The spectral decomposition yields that the largest eigenvalue $\l_{0,h}$
and the relative eigenvectors $\bra{h,0},\ket{h,0}$ are analytic in $z$ and
the denominators $|\braket{h,0}{h\pm1}|$ (see Eq.\equ{e1.4}) are bounded
below by a positive $\d$ and, finally the matrices $\Th=\frac1{2\p i}\oint
\frac{\z d\z}{\z-T)}$ with $T=T_h$ and the integral on a contour
surrounding the spectra of the $T_h$ (or $T(x)$) but excluding $\l_{0,h}$
has also analytic matrix elements uniformly bounded.

Therefore the $I(J,\Bs_J)$ are holomorphic in $|z|<\n'$ and satisfy
essentially the same bounds sufficient for the cluster expansion
convergence, \ie Eq.\equ{e2.5} with suitable $\d,\g$: and $\h$ is small for
large spectral gap.

If the spectral gap of the $T_h$ is not small it is,
nevertheless, smaller than a prefixed $\g$ for the family of matrices
$T'_h=T_h\cdots T_{h+q-1}$ if $q$ is large enough: the analyticity holds,
therefore, as above for the matrices $T'_h$ and consequently for $T_h$.  \*

This concludes a proof of the Theorems.

\def\SEC{Comments}
\section{\SEC}
\label{sec3}\iniz

Neither the theorem in \cite{FK960}, nor its extensions in \cite{Ra979},
have been used, the ergodic theorem being sufficient in the simple cases
considered. The general and deep result in \cite{FK960} does not give 
analyticity: for analyticity a more restricted class of matrices has to be
considered, {\it e.g.} the class considered in \cite{Ru979} or here.

A simple application of theorem 1 is to matrices $(T_n)_{\s,\s'}>0$ with
$1\le\frac{\max_{\s,\s'}(T_n)_{\s,\s'}}{\min_{\s,\s'}(T_n)_{\s,\s'}}\le
C<\infty$: they have the cone property with $\G$ the
cone of the vectors with components $\ge0$ and some, $n$-independent, cone
${\G\,}'$, \cite[Sec.3]{FK960}. This can be immediately applied to obtain free
energy analyticity for a $1D$ spin glass with short range interaction as
remarked in \cite[p.69]{Ru979}: indeed the positivity of $(T_n)_{\s,\s'}>0$
turns the problem into one in Statistical Mechanics with interaction 
$J_{\s\s'}=\log (T_n)_{\s,\s'}$, \cite[p.121]{Ru978}.

The analysis is fully constructive for what concerns the contents of
Theorem 1. In fact lemma 1 can be replaced by the solution of finitely many
eigenvalue problems, like the determination of the largest eigenvalue of
the matrices $T_1,\ldots,T_{N_0}$ or their products with $N_0$ that can be
computed {\it a priori}, if the approximation needed is given.
$N_0$ is directly related to the maximum size of the
polymers necessary to achieve a desired approximation: it is {\it a
  priori} determinable through the value of $\a$ appearing in lemma 2, the
estimate Eq.\equ{e2.4} and the cluster expansion estimates.

Constructivity is only lost, as usual, in the application of the ergodic
theorem, as there is no control on which is the set of matrices for which
the limits like \equ{e2.14}, exist.

Nevertheless in the case of sequences of matrices
randomly chosen with respect to a $\t$-ergodic measure the
determination of the {\it maximal Lyapunov exponent} (or minimal, depending
on the cone property holding)) can be again expressed constructively, as
the integral of the function appearing inside the sum in Eq.\equ{e2.14}
setting $j=1$, in which the first term is explictly known while the second,
\ie $\F(1)$, can be expressed to any prefixed accuracy by the cluster
expansion.

The theorems are not optimal: for instance invertibility of the matrices
$T_j$, absent in \cite{Ru979,Ra979}, is used only in item (3) of theorem 1:
for the remaining statements it is not needed.

The point of this work has been to show how the cluster expansion
technique can be of great help in problems that can be cast into a
statistical mechanics context: after all it has been among the major
achievements in equilibrium statistical mechanics of the XX century. Its
use is limited to special problems but when applicable (as here) it
gives a complete and constructive solution.

 \appendix


\centerline{\Large\bf Appendices}

\def\SEC{Algebraic properties of cones}
\section{\SEC}
\label{appA}\inizA

Call $\e$, ``inclination'', the minimum angle between pairs of vectors in
$\G$ and $\G'$; $\th$, ``opening'', the maximum angle betwee pairs of
vectors in $\G$ and $\th'$ the maximum angle between vectors in $\G'$:
$\p>\th>\th'>0,\e>0$.

\0{\it proof of Lemma 1:} (following \cite{Ru979}) 
Let $T$ be a $d\times d$ matrix and let $\G,{\G\,}'$ be proper, convex, closed
cones (with apex at the origin $O$) in $R^d$. Suppose that
$T\,\G\subset{\G\,}'$ and a relative inclination $\e>0$ of $\G$ to
${\G\,}'$.  

Let $\G^*=\{w|
\braket{w}{v}\ge0,\, \forall v\in\G\}$.
Then, fixed $0\ne v_0\in \G$ and $0\ne w_0\in\G^*$ the maps
\be \textstyle
v\to \frac{T v}{\braket{T^*w_0}{v}},\quad{\rm and}\quad
w\to \frac{T^* w}{\braket{w}{Tv_0}}\Eq{eA.1}\ee
map continuously the convex compact
sets $\{v| \braket{w_0}{v}=1\}$ and, respectively, $\{w|
\braket{w}{v_0}=1\}$ strictly into themselves.  Hence $\exists$
$a\in\G$ and $a^*\in \G^*$ which are fixed points of the maps,
respectively; hence, if $b\defi\frac{a}{||a||}$
and $b^*\defi\frac{a^*}{||a^*||}$
\be\textstyle
\frac{Tb}{||Tb||}=b,\qquad \frac{T^*b^*}{||T^*b^*||}=b^*\Eq{eA.2}\ee
Let $\lis
  T\x\defi\frac{T\x}{||T b||}$ and let
\be K\defi\{\x| \braket{b^*}{\x}=0,\ {\rm and}\  b+\x\in\G\}\Eq{eA.3}\ee
Since $\frac{T b}{||T b||}=b$ and $\frac{T^* b^*}{||T^* b^*||}=b^*$ the
set $K$ is mapped into itself by $\lis T$ ({\it e.g.} $\braket{b^*}{T\x}=0$
and $b+\frac{T\x}{||T b||}=\frac{T(b+\x)}{||T b||}\in\G$) and since the
cone $\G$ is shrunk by $T$ the set $K$ is mapped into $\lis T K\subset \a
K$ with $\a<1$ (determined by the inclination and opening angles, see
Sec.\ref{sec2}).

Hence $\lis T^n(b+\x)=b+\lis T^n\x$ and $||\lis T^n\x||\le \a^n$. For any
$v\in\G, v\ne0,$ there is a $\n\ne0$ such that $v=\n b+\x$, $\x\in K$, so that
\be\frac{T^n(\n b+\x)}{|| T^n(\n b+\x)||}
\equiv\frac{\lis T^n(\n b+\x)}{||\lis T^n(\n b+\x)||}\equiv 
\frac{(\n b+O(\a^n))}{||\n b+O(\a^n)||}\tende{n\to+\infty} \l b\Eq{eA.4}\ee
because $T$ and $\lis T$ are proportional: notice that the above analysis
implies that the largest eigenvalue $\l_0$ of $T$ is positive and that it
is the unique eigenvalue of $T$ with maximum modulus.  \*

\0{\it proof of Lemma 2:} (following \cite{Ru979}) Let $T'\,\defi\, T_1
T_2\cdots T_p$ and let $\Bth,\Bth^*$ be the, respective, normalized
eigenvectors with maximum modulus eigenvalue $\l'>0$ for $T'$ and $(T')^*$
(existing by lemma 1).  Define
\be\eqalign{
\Bth_{p}\defi&\Bth,\ \Bth_{p-1}=
\frac{T_{p}\Bth_{p}}
{||T_{p}\Bth_{p}||}, \ldots,
\Bth_{0}=
\frac{T_{1}\Bth_{1}}
     {||T_{1}\Bth_{1}||}=\Bth
\cr
\Bth^*_{0}\defi&\Bth^*,\ \Bth_{1}^*=
\frac{T_1^*\Bth^*_{0}}{||T^*_1\Bth^*_{0}||}, \ldots,
\Bth^*_{p}=\frac{T^*_{p}\Bth^*_{p-1}}
     {||T^*_{p}\Bth^*_{p-1}||}=\Bth^*\cr}\Eq{eA.5}\ee
By the argument in the proof of lemma 1 the action of $T_j$ on the plane
orthogonal to $\Bth^*_{j}$ maps it on the plane $\Bth^*_{j+1}$ and
contracts by at least $\a<1$. Therefore $T_1T_2\cdots T_p$ contracts by at
least $\a^p$ in the space orthogonal to $\Bth^*$, proving Lemma 2.  \*

\def\SEC{Cluster expansion: a rehearsal}
\section{\SEC}
\label{appB}\inizA

This section follows \cite[Ch.7]{GBG004} (in turn based on
\cite{Ru964,GMM973}) and it is here only for the purpose of making the
paper self-contained for the reader.  Cluster expansion is an algorithm to
compute the logarithm of a sum
\be\X=\sum_{\V J}^* \z(\V J)\equiv\sum_{\V J}^*\prod_i \z(J_i)^{n_i}
\Eq{eB.1}\ee
where: (1) $\V J=(J_1^{n_1},\ldots,J_\NN^{n_\NN})$ with $J_i$'s subsets
in a box $\L$ on a $d$-dimensio\-nal lattice (here $d=1$) called {\it
  polymers} and $n_i\ge0$ are integers defining the ``multiplicity'' of
each (or ``counting'' how many times each set is counted) hence
$\NN=2^{|\L|}$. The sets $J$ could be decorated by associating to each site
$k\in J$ a ``spin'', \ie a variable assuming $d-1$ values. However in the
following the decorations will not be mentioned as they would only make the
notations heavier. In the applications in Sec.\ref{sec2} the decorations
will be necessary and the formulae of this section (which correspond to the
case $d=2$, \ie all spins $1$) are directly usable simply by imagining that
each $J$ is in fact a pair $Y=(J,\Bs_J)$ where $\Bs_J=(\s_j)_{j\in J}$ and
$\s_j=1,\ldots,d-1$.  \\
(2) $\z(\V J)=\prod \z(J_i)^{n_i}$ with $\z(J)$ (small) constants called
  {\it activities}, $\z(\emptyset)\defi1$.
\\ (3) the $*$ means that the sum runs over the $\V J$'s in which no two of
the $J_i\in \V J$ with multiplicity $n_i>0$ overlap in the sense that they
contain pairs of points at distance $\le 1$ on the lattice. If $\wt{\V J}$
denotes the sets in $\V J$ which have positive multiplicity then the $*$
indicates that the sum is restricted to $\V J\equiv \wt{\V J}$ in which
no two of the $J$'s intersect.  

\* 
In applications $\z(J)\ne0$ only for a few of the possible subsets of $\L$.
For instance in the present case $\L$ is the interval $[n,N]$ and the
``polymers'' are just the subintervals.

The $\X$ can certainly be written as $\exp(\sum_{\V J} \f^T(\V J) \z(\V J))$
by expanding the $\log\X$ in powers of the $\z(J)$: of course the sum in
the exponential will involve $\V J$ with $J$'s which can overlap or that
can be counted many times. The $\f^T(\V J)$ are suitable combinatorial
coefficients. 

For instance if $\L$ is just one point $\X=1+z$ can be written as the
exponential of $\sum_{k=1}^\infty \frac{(-1)^{k+1}}k z^k$. If $\L$ consists
of two points, say $1$ and $2$ then the polymers are $\emptyset,1,2,12$ and
$\X=1+z_1+z_2+z_{12}$ is the exponential of $\sum_{k_1+k_2+k_3>0}
\frac{{(-1)^{k_1+k_2+k_3+1}}(k_1+k_2+k_3-1)!}{k_1!  k_2!k_3!} z_1^{k_1}
z_2^{k_2}z_{12}^{k_3}$.

The cluster expansion is the general form of the above examples.  It is of
interest, for instance, if $\sum_{\V J}^\& |\f^T(\V J)| |\z(\V J)|<+\infty$
where the $\&$ means that the sum is restricted to $\V J$'s which contain
any fixed point $x\in\L$ ({\it i.e.} with $x\in\cup_{J\in\wt{\V J}} J$). It
is therefore necessary to determine conditions that imply the mentioned
convergence.

The first step is to define $\V J+\V J'$ simply as $J_1^{n_1+n'_1},\ldots,
J_\NN^{n_\NN+n'_\NN}$, {\it i.e.} as the family of polymers with
multiplicities equal to the sum of the corresponding ones in $\V J$ and $\V
J'$. Let
\be\eqalign{ 
\FF=&\hbox{\rm set of functions $F(\V J)$}\cr
\FF_0= &\hbox{\rm set of functions $F(\V J)$ with $F(\emptyset)=0$}\cr
\FF_1= &\hbox{\rm set of functions $F(\V J)$ with $F(\emptyset)=1$}\cr
\V 1(\V J)=& 
\cases{0 &  if $\V J\ne \emptyset$\cr1& if $\V J=\emptyset$}\cr
f\in&\FF_1 \ \otto \ \wt f\,\defi f -{\bf 1}\in \FF_0\cr
}\Eq{eB.2}\ee
and remark that $f\in \FF_1$ can be written $f={\bf 1}+ \wt f$ with $\wt
f\in\FF_0$.

Then if $f*g(\V J)\defi \sum_{\V J_1+\V J_2=\V J} f(\V J_1)g(\V J_2)$, for
$f,g\in \FF$ define
\be\eqalign{
&{\rm Exp} f(\V J)=\sum_{k=0}^\infty \frac{f^{*k}(\V J)}{k!},\quad
  f\in\FF_0
\cr
& 
{\rm Log} f(\V J)=\sum_{k=1}^\infty 
\frac{(-1)^k \wt f^{*k}(\V J)}{k},
\quad f={\bf 1}+\wt f\in\FF_1
\cr
&f^{*-1}=\sum_{k=1}^\infty (-1)^k \wt f^{*k},\qquad f={\bf 1}+\wt
f\in\FF_1
\cr
&\media{f}=\sum_{\V J\subset\L } f(\V J),\quad f\in\FF
\cr
}\Eq{eB.3}\ee
here all sums over $k$ are finite sums for $f$ in the corresponding domains.

A key remark is %
\be\kern-3mm\eqalign{
&{\rm Log}\,( {\rm Exp}( f))=f\qquad \forall \ f\in\FF_0,\qquad
{\rm Exp}\,({\rm Log} (f))=f\qquad \forall \ f\in\FF_1\cr
&f^{*-1}*f={\bf 1},\quad  \forall\ f={\bf 1}+\wt
f\in\FF_1,\qquad \media{f*g}=\media{f}\media{g}\cr}\Eq{eB.4}\ee
If $\ch(\V J)=\prod \lis\ch(J_i)^{n_i}$ is a {\it multiplitive
  function} $\ch\in\FF$ then $\media{f*g \ch}=\media{f\ch}\media{g\ch}$
so that if $\f\in\FF_1$ and $\lis\ch(J)=\z(J)$

\be\media{f\cdot\z}=\media{{\rm Exp( Log (f\cdot \z)}}=\exp{\media{({\rm
      Log}f\cdot\z)}}\Eq{eB.5}
\ee
Therefore call $\V J$ compatible if $n_i=0,1$ ({\it i.e.} $\V J=\wt{\V J}$)
and the elements of $\wt {\V J}$ are not connected then if 

\be\f(\V J)=\cases{0& if $\V J$ is not compatible\cr1&
otherwise\cr}\Eq{eB.6}\ee
then $\f\in\FF_1$ and $\f^T={\rm Log} \f\in \FF_0$ makes sense and

\be \X=\media{\f\cdot\z}=\exp\media{\f^T\cdot \z}
\equiv \exp\sum_{\V J}\f^T(\V J)\,\z(\V J)\Eq{eB.7}\ee
which is the exponential of a power series in the $\z(J)$ variables. 

Calculating $\f^T(\V J)$ requires computing the sum of finitely many
quantities: if $\V J$ is represented as a set of ``points'' or ``nodes'' and
if $G$ is the graph obtained by joining all pairs of polymers in $\V J$
which are ``incompatible'' (regarding as different, and incompatible with
each other, the $n_i$ copies of $J_i$) it is, ({\it e.g.}  see
\cite[Eq.(4.21)]{GMM973}),
\be \f^T(\V J)=\frac1{\prod n_i!}\sum_{C\subset G}^* (-1)^{\# \ of\ edges\ in
  \ C}\Eq{eB.8}\ee
where the $*$ means that the sum is restricted to the subgraphs of $G$ which
visit all polymers in $G$: their number is huge, growing faster than any
power in the number of polymers so that convergence occurs because of
cancellations due to the relation in Eq.\equ{eB.12}.

The series in Eq.\equ{eB.7} is certainly convergent for $\z(J)$'s small
enough: {\it however the radius of convergence might be very small and $\L$
  dependent}.

Define the {\it differentiation operation} as

\be(D_\G\Ps)( {\V H} )\defi \Ps(\G+ {\V H}) \fra{(\G+ {\V H} )!}{ {\V H}
  !}\Eq{eB.9}\ee
with $\G!=\prod_{i=1}^s n_i!$. The name is attributed
because of the validity of the
following rules: 
\be\eqalign{
&D_\g(\Ps_1*\Ps_2)=(D_\g\Ps_1)*\Ps_2+\Ps_1*(D_\g\Ps_2) , \cr
&D_\g{\rm Exp}\Ps=
D_\g\Ps*{\rm Exp}\Ps , \cr
} \Eq{eB.10}\ee
A direct check of the above relations can be reduced to the case in which
$\G=n\g$, {\it i.e.} to the case in which there is only one polymer species
$\g$, and the check is left to the reader.  The first relation above, {\it
  Leibniz rule}, can be seen as a consequence the combinatorial identity
$\sum_{p_1+p_2=n} {{q_1}\choose{p_1}} {{q_2}\choose{p_2}}= {
  {q_1+q_2}\choose{n}}$ for all $n,q_1,q_2$ with $n\le q_1+q_2$.

The definitions lead to the derivation of the expression for $\f^T(\G)$ in
\equ{eB.8}: which not only is quite explicit but also 
implies immediately that $\f^T(\G)$ vanishes for nonconnected $\G$'s.

To determine sufficient conditions for the convergence which are
independent on the size of $\L$ let $\wh\f(\V Y)\defi\f(\V Y)\z(\V Y)$ and
$\D_{\V J}(\V Y)\defi \wh\f^{*-1}*D_{\V J}\wh \f)(\V Y)$. Then if $\g$ is a
polymer, and $\V J,\V Y$ are polymer configurations
\be \eqalign{
\D_{\g +\V J}(\V Y)=&\sum_{\V Y_1+\V Y_2=\V Y}\wh 
\f^{*-1}*(\V Y_1)\f(\g+\V J+\V
Y_2)\z(\g+\V J+\V Y_2)\cr
=&\z(\g)\sum_{\V Y_1+\V Y_2=\V Y}\wh \f^{*-1}(\V Y_1)*\f(\g+\V J+\V
Y_2)\z(\V J+\V Y_2)\cr
\cr}\Eq{eB.11} \ee

\kern-7mm
Here no factorials appear because $\f(\V J)$ vanishes 
unless $\V J=\wt{\V J}$.

\*
Remark that $\f(\g+\V J+\V Y_2)=\f(\V J+\V Y_2)\prod_{\g'\in\V
  Y_2}(1+\ch(\g,\g'))$ with $\ch(\g,\g')=0$ if $\g,\g'$ do not
overlap and $\ch(\g,\g')=-1$ otherwise, so that $\f(\g+\V J+\V Y_2)=\f(\V
J+\V Y_2) \sum^*_{\V S\subset \V Y_2} (-1)^{|\V S|}$, with $|\V S|=$ number
of polymers in $\V S=(s_1,s_2,\ldots)$ and $*$ means that the $s_i$ overlap
with $\g$, for all $i$. Hence setting $\V Y_2=\V S+\V H$
\be\kern-2mm \eqalign{
\D_{\g +\V J}(\V Y)=&
\z(\g)\sum_{\V S\subset\V Y}^*
\sum_{\V Y_1+\V H=\V Y-\V S}\kern-3mm
\wh \f^{-1}(\V Y_1)\f(\V J+\V S+\V
H)\z(\V J+\V S+\V H)
\cr
=&\z(\g) \sum^*_{\V S\subset \V Y}(-1)^{|\V S|} \D_{\V J+\V S}(\V Y-\V S)
\cr}\Eq{eB.12}\ee
Let $r(\g)\ge|\z(\g)|$  and $r(\V X)=\prod_{\g\in \V X} r(\g)$; then
\be I_m\defi\sup_{1\le n\le m}
\sup_{|\V J|=n} \sum_{\V Y, |\V Y|=m-n}\frac{|\D_{\V J}(\V
  Y)|}{r(\V J)}\Eq{eB.13}\ee
and $I_1$ is then $I_1=\sup \frac{|\z(\g)|}{r(\g)}$ and recursively $I_{m+1}
\le \m^m I_1$ where
\be\m\defi \sup_\g \frac{|\z(\g)|}{r(\g)} 
\exp{\sum_{J}^*r(J)}\Eq{eB.14}\ee
where here $J$ is a single polymer (intersecting $\g$): see
\cite[Eq. 7.1.28]{GBG004} for more details on the algebra.  Therefore
$I_{m+1}\le \m^m I_1$, if $\m<1$.

The latter property $\m<1$ holds in various applications, notably in the
present work, to bound $\O(n,N)$ as
well as a few more quantities.

The method has several other applications, see
\cite{GMM973}, \cite[Ch.7]{GBG004}. Here the polymers $J$ will be
$\Bs_J$ corresponding to intervals $J$  (on
the lattice $[1,N]$) with the associated spin structures $\Bs_J$.
We shall make use of Eq.\equ{eB.7}
and, by Eq.\equ{eB.7} and the third in \equ{eB.10}, of
\be \eqalign{
P(J)\defi&\frac{\sum_{\V H\ni J}\z(\V H)}{\X}=
\frac{\media{D_J \f \z}}{\media{\f\z}}=
\media{\wh\f^{*-1}*D_J\wh\f} 
\cr
&=\media{\wh\f^{*-1}*D_JExp(\wh\f^T)}
=\media{D_J\wh\f^T},\qquad \wh\f\equiv\f\z\cr}\Eq{eB.15}\ee
In an ensemble in which the polymer configurations $\V J$ in $\L$ are given
a weight proportional to $\prod_{\g\in\V J}\z(\g)$ this would be the
probability of finding a configuration of polymers containing the polymer
$J$ if $\z(\g)\ge0$. Hence the complementary sum $P'(J)\defi\frac{\sum_{\V H\in
    J}\z(\V H)}{\X}$ will be such that $P(J)+P'(J)=1$.

\def\SEC{Dynamical systems application}
\section{\SEC}
\label{appC}\inizA

Let $\FF$ be a smooth compact manifold and $\t$ a smooth, smoothly
invertible, map on $\FF$ (take smooth to mean $C^\infty$, for simplicity).
At each point $x\in \FF$ there are proper closed convex cones $\G(x)\supset
{\G\,}'(x)$, with apex at $x$ in a linear space $E(x)$ of dimension $d$
smoothly dependent on $x$ (and call its adjoint $E(x)^*$). The cones are
also supposed to depend smoothly on $x$.
\*

\0{\bf Definition:}\label{D2} {\it The minimum angle between vectors
  on the boundary of $\G(x)$ and on that of ${\G\,}'(x)$ will be called
  inclination $\e(x)$; while the maximum angle between vectors in $\G(x)$
  will be called $\th(x)$, likewise define $\th'(x)$.}
\*

\0Let $T(x)$, $x\in\FF$, be an invertible mapping of $E(x)$ onto $E(\t x)$,
and \\
$T(x)$ maps $\G(x)$ into  ${\G\,}'(\t x)\subset \G(\t x)$ with 
${\G\,}'(x)/\{x\}\subset \G(x)^0$ and
\\
$T^{\pm}(x),\G(x),{\G\,}'(x),\e(x),\th(x),\th'(x)$ be
smooth, $\p>\th(x)>\e(x),\th'(x)>0$.
\*

\0Making use of Lemma 1, 2 in Appendix \ref{appA} it will not be restrictive to
suppose that $T(x)$ is ``almost diagonalizable'' in the sense that there exist
$\l_0(x)$, $\ket{x,0}\in E(\t x),\bra{x,0}\in E(x)^*$ smoothly dependent on
$x$ and $\Th(x)$ with norm such that $||\Th(x)||/\l_0(x)$ is smaller than a
prefixed quantity ($x$-uniformly):
\be T(x)=\l_0(x)\ket{x,0}\bra{x,0}+\Th(x)\Eq{eC.1}\ee
Then setting $T_h\defi T(\t^{-h}x)$ and repeating the proof of theorem 1
leads to

\* \0{\bf Theorem 3:} {\it Let $T(x)$ be as above.
 Let $x\to
  v(x)\in {\G\,}'(x), ||v(x)||\equiv1$ be a measurable function
(not necessarily continuous), it is
  \\
(1) There are continuous functions $x\to b(x)\in\G(x)$, $x\to \lis \L(x,p)$
and $x\to\L(x,p), p=0,1,\ldots$, such that, for all $p>0,x\in\FF,v$, exist
the limits
\be \eqalign{
b(x)=&\lim_{N\to\infty} \frac{T(x)
\cdots T(\t ^{-(N-1)}x) v(\t ^{-(N-1)}x)}
{\lis \L_v(x,N)}
\cr
\noalign{\vskip2mm}
{\L(x,p)}=&\lim_{N\to\infty} 
\frac{\lis\L_v(\t^{-p} x,N)}{\lis\L_v(x,N)}>0, 
\cr
}\Eq{eC.2}\ee
(2) The vectors $b(x)$ are eigenvectors for products of $T(\t^{-j} x)$
in the sense
\be
b(\t ^{-p}x)=\,\L(x,p)\, T^{-1}(\t^{-(p-1)}x)\ldots T^{-1}(x)\, b(x)
\Eq{eC.3}\ee
(3) $b(x),\L(x,p)$ are $v$-independent and continuous in $x\in\FF$ and
$\exists B$ such that $B^{-1}<||b(x)||<B$; if $T(x)$ is the Jacobian of
$\t^{-1}$ the unit vector $\frac{b(x)}{||b(x)||}$ will be called the
unstable unit vector, or unstable direction, at $\t x$.  \\
(4) The upper and lower limit values $\ell^\pm(x)$ of 
$\frac1p \log\L(x,p)$ as $p\to\infty$  are
constant along trajectories, \ie $k$-independent if evaluated at $\t^{-k} x$.}

\*
The continuity is an extra property due to the continuity of the terms
appearing in the cluster expansion. Analiticity of $b(x),\L(x,p)$ can be
obtained as in the case of theorem 2 under natural analogous assumptions.

If $x$ is chosen randomly with respect to an invariant measure $\r$
then the limits in item (4) are a.e. equal (as in the
case of Sec.\ref{sec2}: via the cluster expansion, they are represented
as ``Birhoff averages''. If $\r$ is ergodic the limits not
only exist but are $x$-independent a.e. and $b(x)$ identifies
the unstable direction at $x$ while $\ell=\ell^+=\ell^-$ is the
maximum Lyapunov exponent.

\vfill\eject
\def\SEC{References}
\bibliographystyle{plain}



\end{document}